\documentclass[preprint2]{aastex}

\usepackage[]{hyperref}
\usepackage[greek,english]{babel}
\usepackage{longtable}
\usepackage{alltt} 
\usepackage{booktabs} 
\usepackage{overpic} 
\usepackage{xcolor}
\usepackage{seqsplit} 
\usepackage{url}

\newcommand{\Sersic}{S\'{e}rsic}
\newcommand{\corser}{core-S\'{e}rsic}
\newcommand{\chisq}{$\chi^{2}$}
\newcommand{\GALFIT}{\textsc{Galfit}}
\newcommand{\CORSAIR}{\textsc{Galfit-Corsair}}

\shorttitle{G{\small ALFIT}-C{\small ORSAIR}: implementing the \corser{} model into G{\small ALFIT}}
\shortauthors{Paolo Bonfini}

\begin{document}
\sloppy 

\title{G{\small ALFIT}-C{\small ORSAIR}: implementing the \corser{} model into G{\small ALFIT}}

\author{Paolo Bonfini}
\affil{Centre for Astrophysics and Supercomputing, Swinburne University of Technology\\ 
       Hawthorn, Victoria 3122, Australia}
\email{pbonfini@swin.edu.au}

\date{
 \begin{center}
  \emph{Accepted for publication in PASP, August 26, 2014}
 \end{center}
}

\begin{abstract}
 \noindent
 We introduce \CORSAIR{}: a publicly available, fully retro-compatible
 modification of the 2D fitting software \GALFIT{} (v.3) which adds an
 implementation of the \corser{} model.

 \noindent
 We demonstrate the software by fitting the images of \mbox{NGC 5557} and
 \mbox{NGC 5813}, which have been previously identified as \corser{} galaxies by
 their 1D radial light profiles.
 These two examples are representative of different dust obscuration conditions,
 and of bulge/disk decomposition.
 To perform the analysis, we obtained deep Hubble Legacy Archive (HLA) mosaics
 in the F555W filter ($\sim${$V$}-band).
 We successfully reproduce the results of the previous 1D analysis,
 modulo the intrinsic differences between the 1D and the 2D fitting
 procedures.

 \noindent
 The code and the analysis procedure described here have been developed for
 the first coherent 2D analysis of a sample of \corser{} galaxies, which will be
 presented in a forth-coming paper.
 As the 2D analysis provides better constraining on multi-component fitting,
 and is fully seeing-corrected, it will yield complementary constraints on the
 missing mass in depleted galaxy cores.
\end{abstract}




\section[Introduction]{Introduction}
\label{Introduction}

\subsection[corser galaxies]{Core-S\'{e}rsic galaxies}
\label{corser galaxies}

The presence of cores in luminous early type galaxies, and their correlation
with the global galaxy properties, has been known since the early ground-based
studies
\citep[e.g.][and references therein]{king:1966,king:1972,king:1978,young,duncan,begelman,kormendy:1985,lauer:1985}.
Through the advent of the high resolution offered by HST it was
possible to characterize the radial profiles of cores.
In particular, it was observed that cores could be approximated by a power law
$I(r)\propto r^{\gamma}$ \citep[e.g.][]{lauer:1991,crane,ferrarese,kormendy:1994,forbes}.

In the same framework, it was also observed that elliptical galaxies and bulges of
disk galaxies (hereafter, commonly labelled as ``spheroids'') present a dichotomy
in the distribution of the power-law index \citep[e.g.][]{jaffe,ferrarese}.
According to this bimodality, galaxies were classified into two sharp categories:
``core'' ($\gamma <$ 0.3) galaxies, and ``power-law'' ($\gamma>$ 0.5) galaxies,
although later studies suggested the existence of intermediate objects
\citep[e.g.][see also Figure 8 from \citealt{graham:intermediate_gamma}]{ravindranath,rest}.

In order to coherently connect the core/power-law region to the overall shape
of the galaxy profile, \cite{kormendy:1994}, \cite{grillmair:nuker}, and
\cite{lauer:nuker} introduced the ``Nuker law'' (also know as ``Nuker model'').
Similarly to the double power-law model of \cite{ferrarese}, the Nuker law is
composed of two connected power laws, with the difference that the ``smoothness''
of the transition is regulated by an additional parameter (\emph{$\alpha$}).
This 5-parameter model has the same functional form
as the double power-law density model of \citet[his Equation 43]{hernquist},
but it is applied to the projected surface brightness rather than to the internal
density profile.
The Nuker model soon became the reference tool for the study of the cores of early
type galaxies.

Later studies, however, demonstrated that the Nuker model, and hence the relevant
classification scheme, are critically affected by the observational and fit
conditions.
For example, \cite{faber} and \cite{rest} showed that a ``core'' galaxy can be
misclassified as ``power-law'' if the break radius is below the resolution limit,
or if the transition is so smooth that the index $\gamma$ is overestimated
because of poor sampling of the innermost radii.
In addition, \cite{graham:corser} showed that a Nuker fit is highly
dependent on the radial extent of the fit.

In order to overcome these limitations,
\cite{graham:corser} and \cite{trujillo:corser}
introduced the ``\corser{}'' model,
which represents a smooth connection between a \cite{sersic}
component (descriptive of the outer spheroid profile), and a single power-law
component (descriptive of the core).
These authors showed that the spheroids of the
``core'' and ``power-law galaxies'' could have their full radial extent fit simply
by a \corser{} and a \Sersic{} model, respectively.
 
Cores are believed to be created by the scouring action of a binary
supermassive black hole (SMBH) system produced in ``dry'' major merging events,
which ejects stars via a three-body interaction.
This hypothesis was first formulated by \cite{begelman}, and its plausibility was
later confirmed by numerical simulations \citep[e.g.][]{ebisuzaki,merritt}.
In this sense, a \corser{} galaxy is simply a \Sersic{} galaxy with a
depleted core.

In the last decade, the sample of know \corser{} galaxies (identified by
their 1D radial light profile) has been constantly expanded by a collection of studies
focused on a critical review of the structure of early type galaxies
\citep[e.g.][]{trujillo:corser,ferrarese:2006}, and/or  aimed to the study of the
relations between SMBHs and their host galaxies
\citep[e.g.][]{richings,dullo:2012,dullo:2013a,dullo:2013b}.
In addition to these samples, possible \corser{} candidate galaxies may be ``hidden''
in the sample by \cite{lauer:2007}, who gathered information from diverse studies
which used the Nuker law to characterize the inner light profiles of galaxies.
 
\subsection[2D modeling of core-Sersic galaxies]{2D modeling of \corser{} galaxies}
\label{2D modeling of corser galaxies}

As mentioned above, coherent studies of \corser{} galaxies have been
based solely on 1D radial light profiles.
The 1D fitting procedures have the remarkable advantage of being able to work
in poor $S/N$ conditions.
For example, in the isophote fitting method \citep{ellipse} - which is arguably
the most widely used - higher $S/N$ is provided because each point on the radial
profile is calculated as azimuthally-averaged surface brightness along a curve.
On the other hand, in 1D methods the definition of ``radial'' profile itself
can be ambiguous.
In isophote fitting, isophotes at different galactocentric
radii can have different ellipticities and position angles, so that the radial
intensity profile is actually extracted along the curved path defined by the
apocenters of the ellipses.
In major/minor axis fitting, the presence of multiple galaxy components with
intrinsically different orientations make the definition of ``major'' or ``minor''
axis arbitrary; moreover, the true major axis could be along the line of sight.
An additive limitation is that 1D models are intrinsically assumed to be
axisymmetric.

On the other hand, 2D models with fixed ellipticity and position
angle cannot accommodate such changes, whereas isophote fitting does
so naturally.
However, given the often very different ellipticity and position angle of
bulges and disks, fitting in 2D can help disentangling those components in
the projected galaxy image, hence releasing part of the degeneracy in the parameters.
These issues and the advantages of 2D fitting have been discussed extensively
in the literature \citep[see e.g.][and references therein]{ravindranath,BUDDA}. 

There exists a number of public softwares performing multi-component 2D fit
of galaxies, offering a variety of features (e.g. analytic models), and tuned to tackle
different objects or galaxy characteristics (e.g. distant/nearby galaxies,
symmetric/asymmetric features, etc.), such as \textsc{Gim}2D \citep{GIM2D},
\textsc{Budda} \citep{BUDDA}, \textsc{Budda v.2} \citep{BUDDA2}, and \GALFIT{} \citep{GALFIT}.

In particular, \textsc{Gim}2D and \GALFIT{} are arguably the codes which have received
the biggest attention from the astronomical community, and have undergone the most
extensive testing.
In their review, \cite{haussler} discuss the treatment of the uncertainties
in the two codes, and their robustness in crowded-field environments, and conclude
that \GALFIT{} has an advantage in terms of dealing with contaminating objects
(which can be fit independently), although both codes are found to underestimate
true uncertainties.

More than the other available codes, \GALFIT{} was updated along the years to include
new features: the latest update of the software (v3\footnote{\noindent
 \href{http://users.obs.carnegiescience.edu/peng/work/galfit/galfit.html}{\seqsplit{http://users.obs.carnegiescience.edu/peng/work/galfit/galfit.html}}
};
2010) implemented the possibility of fitting complex and asymmetric galaxy features such as spiral arms, rings, bars, etc.
via Fourier modification of standard models (exponential disk, \Sersic{}, etc.) or
coordinates rotations along the radial direction.
Recently, a plethora of codes based on \GALFIT{} were developed in order to automatize
the source detection and fit \citep[e.g. \textsc{Galapagos};][]{GALAPAGOS}, or to expand its
capabilities (e.g. \textsc{Galfit}M\footnote{
 \url{http://nottingham.ac.uk/astronomy/megamorph/}
} allows simultaneous multi-band fitting); \GALFIT{} also represents the basic
ingredient for some software suites designed for the decomposition of galaxy images,
such as \textsc{MegaMorph}\footnote{
 \url{http://nottingham.ac.uk/astronomy/megamorph/}
}.

With \CORSAIR{}, we expanded further the potential of \GALFIT{} v.3
by including the \corser{} profile, which was not present among its native models.

To our knowledge, there is no other 2D fitting code which is natively able to fit a
\corser{} profile, except for Imfit\footnote{
 \url{http://mpe.mpg.de/~erwin/code/imfit/}
}
(Erwin 2014; in preparation), which we recently learnt has simultaneously been under
development.
Comparing the performance of the two codes is beyond the scope of this article.

In the following, we introduce our implementation of the \corser{} model
into \GALFIT{}, and we describe its potential and limitations.
In a forth-coming paper (Bonfini et al. 2014, in preparation), we will present a study
of a large sample of \corser{} galaxies based on this software.
This article is structured as follows.
In \S \ref{Implementation of the corser model in GALFIT}
we briefly present how \GALFIT{} works, and we give the definition of the \corser{} model, along with the basic instructions for its usage. 
In \S \ref{A practical example: analysis of HST data} we show a practical application of the software to
the HST data of two \corser{} galaxies: \mbox{NGC 5557} and \mbox{NGC 5813}.
We evaluate the reliability of the results, and we report some troubleshooting
to assist the reader interested in using our model.
Our conclusions are summarized in \S \ref{Summary and conclusions}.
We also provide an appendix focused on the details of the examined galaxies
(Appendix \ref{Notes on sample galaxies}), and finally an appendix summarizing
the instructions on how to use our code (Appendix \ref{CORSAIR how to})

\section[Implementation of the corser model in GALFIT]{Implementation of the \corser{} model in GALFIT}
\label{Implementation of the corser model in GALFIT}

\GALFIT{} is a 2D-fitting software coded in {\textsc c} and based on the
Levenberg-Marquardt method \citep[e.g.][]{bevington} for the minimization of the
\chisq, which is defined in the classical way, i.e.:

\begin{equation}
 \chi^{2} = \displaystyle\sum_{x=1}^{nx} \displaystyle\sum_{y=1}^{ny} { [f_{data}(x,y) - f_{model}(x,y)]^2 \over \sigma(x,y)^2}
 \label{equation:GALFIT_Chi^2} 
\end{equation}

\noindent
where $f_{data}(x,y)$ is the value of the ($x$,$y$) pixel of the input image,
$f_{model}(x,y)$ is the value of the corresponding pixel in the PSF-convolved 
model image generated at each iteration,  $x$ and $y$ refer to the pixels in the
images, and $\sigma(x,y)$ is the ``sigma'' image.
The sigma image is used to weight the discrepancy of the model to the data,
given the statistical error of the observed counts at each pixel position.

In order to determine in which direction to look for the next iteration parameter
set, the Levenberg-Marquardt method requires knowledge of the local curvature of
the \chisq{} space, which is given by the Jacobian (or ``convolution'') matrix
(i.e., the matrix of the partial derivatives of the model with respect to its
parameters).
When the global minimum is found, the parameter uncertainties are obtained from
the final convolution matrix, and represent the 1-$\sigma$ errors under
the assumption that the data probability distribution function is Gaussian, and
that the \chisq{} space around the minimum has a parabolic shape.

In order to include the \corser{} profile in \GALFIT{}, we had to code
the model and its partial derivatives.
The model surface brightness profile is defined as\footnote{
 Notice that in this formulation we have chosen a specific meaning for $b_{n}$:
 see discussion in the text
}:

\begin{small}
\begin{equation}
 I(r) = I'\left[1+\left({r_{b} \over r}\right)^{\alpha}\right]^{\gamma / \alpha}
        \exp \left[ -b_{n}\left({ {r^{\alpha} + r_{b}^{\alpha}} \over r_{e}^{\alpha} }\right)^{{1 / (n\alpha) }}\right]
 \label{equation:corser}
\end{equation}
\end{small}  

where:

\begin{small}
\begin{equation}
 I' = I_{b} 2^{-\gamma/\alpha} \exp \left[b_{n} 2^{1 / n\alpha} (r_{b}/r_{e})^{1/n}\right]
 \label{equation:IP}
\end{equation}
\end{small}  

The model represents a smooth connection between a \Sersic{} component (descriptive
of the overall spheroid profile), and a power-law component (descriptive of the ``core''),
separated at a characteristic \emph{break radius} $r_{b}$.
The transition is modulated by the parameter $\alpha$, which determines the sharpness
of the changeover.
It is trivial to show that far from $r_b$, the second term of Equation
\ref{equation:corser} represents a power-law component with slope -$\gamma$, while
the third term represents the classic \Sersic{} component.
The normalization factor $I_{b}$ represents the surface brightness at the break radius.
In this formulation of the \corser{} profile, the other parameters retain the meaning
they have in the classic definition of the \Sersic{} model.
In particular, $b_{n}$ is the normalization factor which lets the parameter $r_{e}$
assume the role of the \emph{effective radius} (half-light radius) of the
\emph{\Sersic{} part} of the model, and is defined by the equation:
\begin{small}
\begin{equation}
 \Gamma(2n) =  2\gamma(2n,b_{n})
 \label{equation:b_n}
\end{equation}
\end{small} 

\noindent
where $\Gamma$ and $\gamma$ represent here the complete and incomplete gamma functions,
respectively \citep[e.g.][and references therein]{graham:sersic}.
An approximated expression for $b_{n}$ was provided by \cite{capaccioli} for the range
0.5 $<$ $n$ $<$ 10, in the form:
\begin{small}
\begin{equation}
 b_{n} = 1.9992n - 0.3271.
 \label{equation:b_n_approximation}
\end{equation}
\end{small} 

For the sake of completeness, we notice that the couple ($b_{n}$,$r_{e}$) can
be replaced - without changing the functional shape - with the couple ($b$,$r_{cs}$),
with $r_{cs}$ now being the half-light radius of the \emph{whole} model, and $b$
defined accordingly \citep[see][Appendix A]{trujillo:corser}.
Though, we chose to implement the definition implied by Equation \ref{equation:b_n},
because it readily allows for a comparison against a \Sersic{} fit.
The two effective radii are anyway very close, as the amount of ``missing light''
is typically less than 1\%
\citep[e.g.][]{graham:mass_deficits,ferrarese:2006,cote,dullo:2013b}.
In our definition of the model, $b_{n}$ is calculated according to Equation
\ref{equation:b_n_approximation} for $n$ $>$ 5, and through a numerical evaluation
of Equation \ref{equation:b_n} otherwise; this is consistent with what happens for
the existing \Sersic{} model in \GALFIT{}.

In practice, for values of $\alpha$ larger than $\sim$10, the transition is already
almost indistinguishable from a discontinuity, therefore we could restrict the code
from sampling an unnecessary large parameter range.
Moreover, the analytical form of the model
and of its derivatives contains several powers of $\alpha$, and it is expected
that the code may encounter numerical
problems if $\alpha$ was to became too large. 
For this reasons, we decided to set a hard constraint on the upper limit of that
parameter, so that $\alpha$ is limited in \CORSAIR{} as:
 0 $<$ $\alpha$ $<$ 20.
The upper limit was selected after extensive testing, in which we observed that
if at any iteration $\alpha$ was going past $\sim$20, then the fit would inevitably
end up diverging.
See \S\ref{Assessing results and troubleshooting} for further troubleshooting
related to this parameter.
Similarly, the power-law slope $\gamma$ has been limited to a physically viable
lower limit of -0.2.
The choice of the aforementioned constraints have been chosen based on the available
data for confirmed \corser{} galaxies \citep{dullo:2013b}.

\medskip
A pre-compiled version of \GALFIT{} including the \corser{} model is available
at the address:

\medskip
\url{http://astronomy.swin.edu.au/~pbonfini/galfit-corsair/}
\medskip

\noindent
along with an example input file, and usage notes, as summarized in Appendix
\ref{CORSAIR how to}.

\section[A practical example: analysis of HST data]{A practical example: analysis of HST data}
\label{A practical example: analysis of HST data}

\subsection[Data set]{Data set}
\label{Data set}

We present here a practical application of our code on two \corser{}
galaxies: \mbox{NGC 5557} and \mbox{NGC 5813}, selected from the catalogue of
``secure'' core galaxies of \citet{dullo:2013b}.
These objects have been chosen because of their relatively large break radii
($r_{b}$), which best suits the illustrative purposes of this example.
Moreover, \mbox{NGC 5813} also includes a disk component
(apart from the \corser{} bulge component), hence it represented a good test for
the robustness of our code with respect to multiple component analysis.
In a following paper (Bonfini et al. 2014, in preparation), we will present
our 2D analysis for the full sample of the \corser{} galaxies of
\cite{dullo:2013b}.

To perform our analysis, we decided to take advantage of the high resolution
offered by HST over large area detectors.
In particular, high resolution is critical for the purposes of the 
current exercise, as previous 1D analysis
have shown that depleted cores have linear sizes of few hundreds parsecs
at most \citep{dullo:2012,dullo:2013b}, which at the distances of our
sample galaxies \mbox{NGC 5557} \citep[47.9~Mpc;][]{cantiello}
and \mbox{NGC 5813} \citep[32.2~Mpc;][]{tonry} translates to few arcseconds.

We profited of the data by the Hubble Legacy Archive (HLA\footnote{
 \url{http://hla.stsci.edu/}
}) project, which offers high
level archival products from the Hubble Space Telescope.
We searched for images in the F555W filter
(similar to the Johnson-Cousin $V$-band), due to its higher $S/N$, and for
the purpose of comparison against literature data \citep[mostly][]{dullo:2013b}.
For both galaxies, the best available data resulted to be single-visit combined
images (``level 2'') for the WFPC2 camera.

These products consist of mosaics of multiple WFPC2 images created via the 
\emph{MultiDrizzle} software \citep{multidrizzle}, [up]binned to the Wide Field
Camera CCD (WF) pixel scale (nominally, 0.0996\arcsec/pixel), aligned north up, and
astrometrically corrected.
The sky level was automatically estimated by \emph{MultiDrizzle} via sigma-clipping
algorithms, and then \emph{subtracted} from the images.
Notice that the image data are expressed in electrons/second, while the standard WFPC2
pipeline products are expressed in DN.
Each mosaic is accompanied by a weight and an exposure image (as produced by 
\emph{MultiDrizzle}), but unfortunately not a bad pixel mask, at least for the
WFPC2 camera.
The details of the sample images are summarized in Table \ref{table:sample_images}.

The HLA also provides combined mosaics of the sole PC chip of WFPC2 with a higher
pixel scale (nominally 0.050$\arcsec$/pixel), which would intuitively appear more
beneficial to our analysis.
Though, as \cite{graham:corser} suggest, in order to constrain the degenerate
$r_{e}$ and $n$ parameters of the model, it is better to extend the \corser{} fit
up to ${\sim}r_{e}$.
Given this warning, for the illustrative purposes of this example we opted for the
whole WFPC2 images (as opposed to the sole PC chip), even if that meant
sacrificing pixel resolution in the core regions.
We will show that - at least for the galaxies under examination - the WF
pixel scale was still sufficient to trace the power-law part of the \corser{} model.

\subsection[Data preparation and fitting]{Data preparation and fitting}
\label{Data preparation and fitting}

The $\chi^{2}$ minimization algorithm of \GALFIT{} requires the user
to provide an initial point from where to start mapping the $\chi^{2}$ space
determined by the fit parameters; in particular, a critical parameter is represented
by the sky level estimate.
\GALFIT{} also allows the user to supply meta-images - such as the sigma (error)
image, or the bad pixel mask - to assist the fitting procedure.
In the following, we detail the necessary preparatory steps tuned to fit
the HLA mosaics, and then proceed to illustrate the fit results in
\S\ref{Fit}.

\subsubsection[Sky level estimation, sigma image, and fit extent]{Sky level estimation, sigma image, and fit extent}
\label{Sky level estimation, sigma image, and fit extent}

The information about the sky level can be input to \GALFIT{} for two purposes:
as a component of the fit model, and in order to calculate the sigma image.

We will focus first on the second aspect.
In our procedure, we let \GALFIT{} generate the sigma image by its internal
calculations.
This image, which in \GALFIT{} represents the Poissonian approximation to the
1-$\sigma$ error at each pixel position (assuming a Gaussian distribution of the image
data), is internally calculated as \citep[][Equation 33]{GALFIT}:

\begin{small}
\begin{equation}
 \sigma(x,y) =  \sqrt{(f_{obj}(x,y) - <sky>)/GAIN + \sigma_{sky}(x,y)}
 \label{equation:sigma}
\end{equation}
\end{small} 

\noindent
where $f_{obj}(x,y)$ is the input image, $<$$sky$$>$ is the median value of the sky,
$GAIN$ is actually the effective gain (i.e., analog to digital gain times the number
of combined images; $NCOMBINE$), and
$\sigma_{sky}(x,y)$ is the sky root mean square (RMS) measured by \GALFIT{}
(which is here assumed to also include the systematic readout noise).

\GALFIT{} calculates the sky level and the RMS via a sigma-clipping
algorithm, after smoothing the image with a Gaussian filter with a hard-coded
FWHM of 2 pixels, and kernel size of 5 pixels.
Obviously, for this procedure to work properly, it is necessary that the
fit area is devoid of sources for the most part: at least 60\% \citep{GALFIT}.
In our images, the wings of the galaxy light was clearly extending well beyond
the aforementioned limit, therefore we had to supply the sky level and RMS to \GALFIT{}
from an independent calculation.
Moreover, as described in \S\ref{Data set}, the HLA products are provided sky-subtracted,
and the sky is evaluated along the HLA pipeline (nominally, by \emph{MultiDrizzle}) in a
similar fashion as for \GALFIT{}, hence it was affected by the same issue, and turned
out to be unreliable.
In particular, both looking at the extended light profiles of the target galaxy, and by
noticing a systematic abundance of negative-valued pixels at the image borders, we
inferred that the sky level was over-estimated due to the contribution of the
galaxy light.

At a first attempt, we tried to restore the sky level by using the relevant information
stored in the \textsc{mdrizsky} keyword within the file headers.
As mentioned in \S\ref{Data set}, the units of the HLA output differ from those of the
standard WFPC2 pipeline.
We warn the reader that the units of \textsc{mdrizsky} are supposed to be those
indicated by \textsc{bunit}, but this header keyword is \emph{not} updated by the
HLA pipeline: the correct units for WFPC2 \textsc{mdrizsky} are DN
(STScI Archive; personal communication).
We therefore added back the \textsc{mdrizsky} value to the images (after proper unit conversion), but the derived sky surface brightness resulted a few magnitudes off
what we expected (i.e. $\sim$22.9 mag/arcsec$^{2}$ for a $V$-band filter as F555W).
We also tried to add back the \textsc{mdrizsky} value to the single native WFPC2 images
(i.e. the *\_c0m.fits) without being able to reproduce the sky counts in the
corresponding drizzled single images (hst\_*\_drz.fits)\footnote{
 Both kinds of images are accessible by using the "more" option under the Image display
 of the HLA.
}.
We deduced that there is a mismatch in the header keyword units, and we discarded
this approach to the sky calculation.

Instead, we estimated the background level and RMS from Poissonian calculations,
using tabulated values for the expected sky brightness in the F555W filter, and the
detector readout noise and dark current characterized by the STScI team, as
described in the following.

The average background value per pixel ($B$) is dominated by the sky noise:
the readout noise is negligible, while the dark current is in principle subtracted
by the HLA pipeline (although the value reported in the \textsc{darkcorr} keyword
suggests it is underestimated).
Hence, $B$ can be expressed simply as:

\begin{small}
\begin{equation}
 B~[DN/pixel] =  {R^{F555W}_{sky} \times EXPTIME \over NCOMBINE \times GAIN}
 \label{equation:B}
\end{equation}
\end{small}

\noindent
where $R^{F555W}_{sky}$ (electrons/pixel) is the sky count rate per pixel expected
for a sky brightness of 22.9 mag/arcsec$^{2}$ observed in the F555W filter, which we
obtained from the WFPC2 Instrument Handbook
\citep[0.052 electrons/second/pixel;][Table 6.4]{WFPC2}, where we used WF chip values as
the galaxies are centered on the PC chip.

On the other hand, the average background RMS per pixel ($RMS_{B}$) can be expressed
by the sum in quadrature of the noise due to the sky ($RMS_{sky}$), the readout noise
($RMS_{rd}$), and the detector dark current ($RMS_{det}$):

\begin{small}
\begin{equation}
 RMS_{B} = \sqrt{RMS_{sky}^2 + RMS_{rd}^2 + RMS_{det}^2}
 \label{equation:RMS_B}
\end{equation}
\end{small}

\noindent
where: 

\begin{small}
\begin{equation}
 RMS_{sky}~[DN/pixel] = {\sqrt{R^{F555W}_{sky} \times EXPTIME} \over NCOMBINE \times GAIN}.
 \label{equation:RMS_sky}
\end{equation}
\end{small}

\noindent
$RMS_{det}$ can be estimated by the Poissonian error on the dark current
rate ($R_{det}$):

\begin{small}
 \begin{equation}
  RMS_{det}~[DN/pixel] = {\sqrt{R_{det} \times EXPTIME} \over NCOMBINE \times GAIN}.
  \label{equation:RMS_det}
 \end{equation}
\end{small}

\noindent
We obtained $R_{det}$ from the WFPC2 Instrument Handbook (Figure 4.8),
where we conservatively adopted the maximum ever recorded value among the WF chips
($R_{det}$ $\sim$ 30~electrons/second/pixel).
Similarly, the readout noise for the combined image can be estimated by applying
error propagation on the readout RMS of each single field ($RD$):

\begin{small}
 \begin{equation}
  RMS_{rd}~[DN] = {RD \over \sqrt{NCOMBINE} \times GAIN}.
  \label{equation:RMS_rd}
 \end{equation}
\end{small}

$RD$ was obtained from the WFPC2 Instrument Handbook (Table 4.2), where we
adopted the maximum readout noise among the WF chips ($RD$ $\sim$ 5.51~electrons).
 
Equations \ref{equation:B} through \ref{equation:RMS_rd} are expressed in terms of
DN as these are the input units required by \GALFIT{}.
The background level and RMS information was passed to \GALFIT{} to generate the sigma
image as per Equation \ref{equation:sigma}.

\begin{figure*}
 \makebox[\linewidth]{
  \begin{overpic}[width=0.48\textwidth]
   {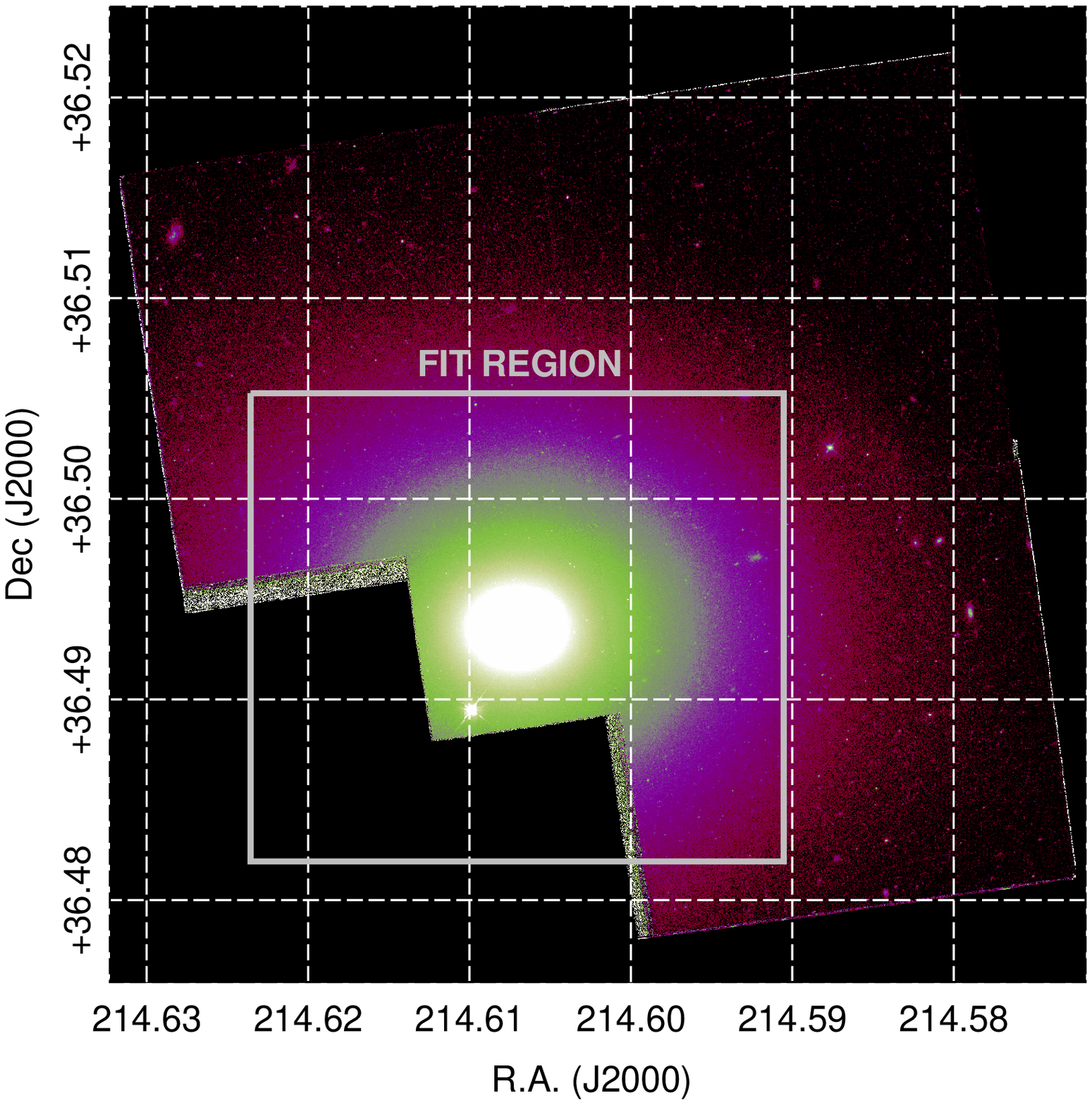}
   \put(20,75){\textcolor{white}{NGC 5557}}
  \end{overpic}
  \includegraphics[width=0.48\textwidth,angle=0]{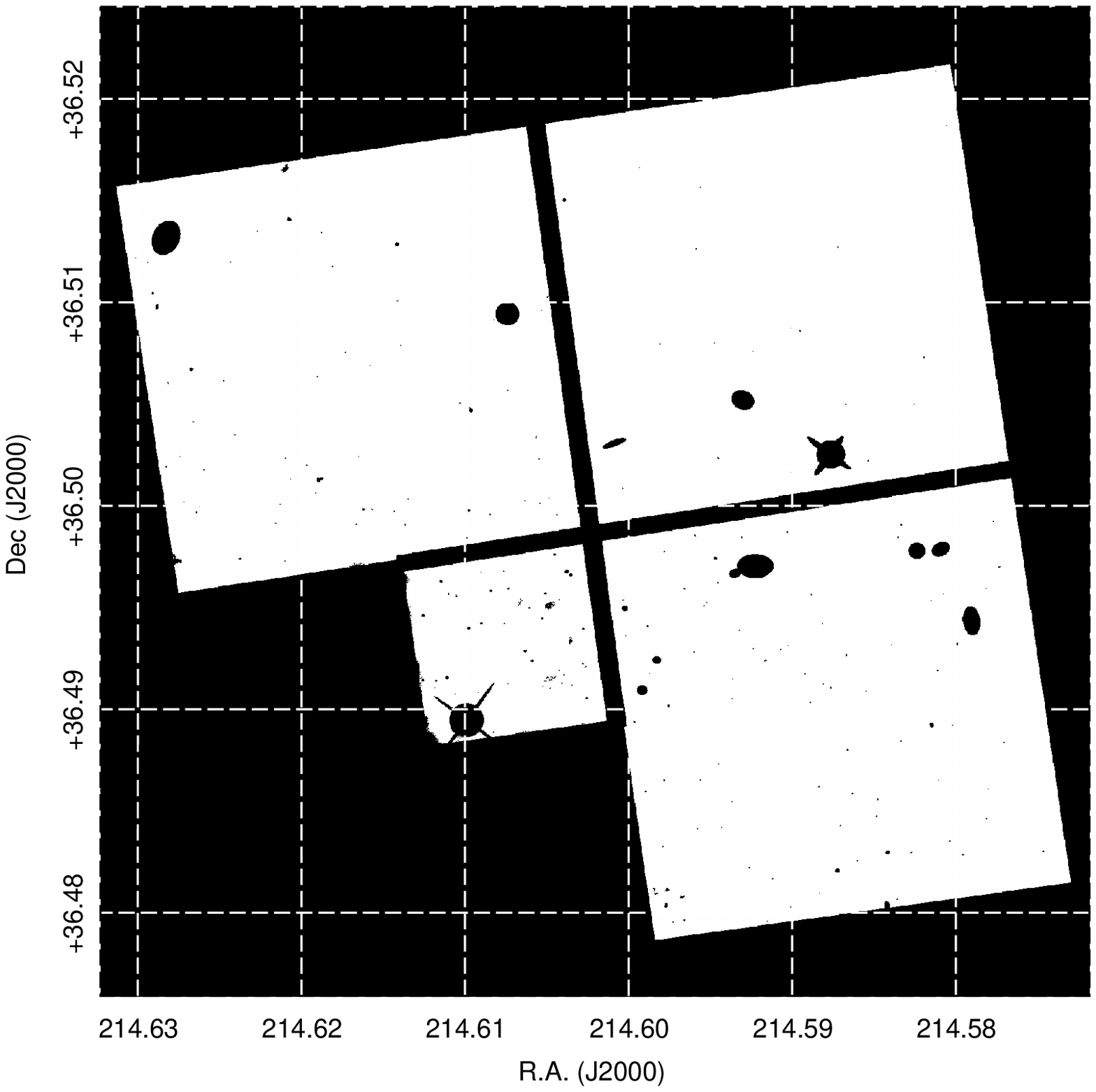}
 }
 \makebox[\linewidth]{
  \begin{overpic}[width=0.48\textwidth]
   {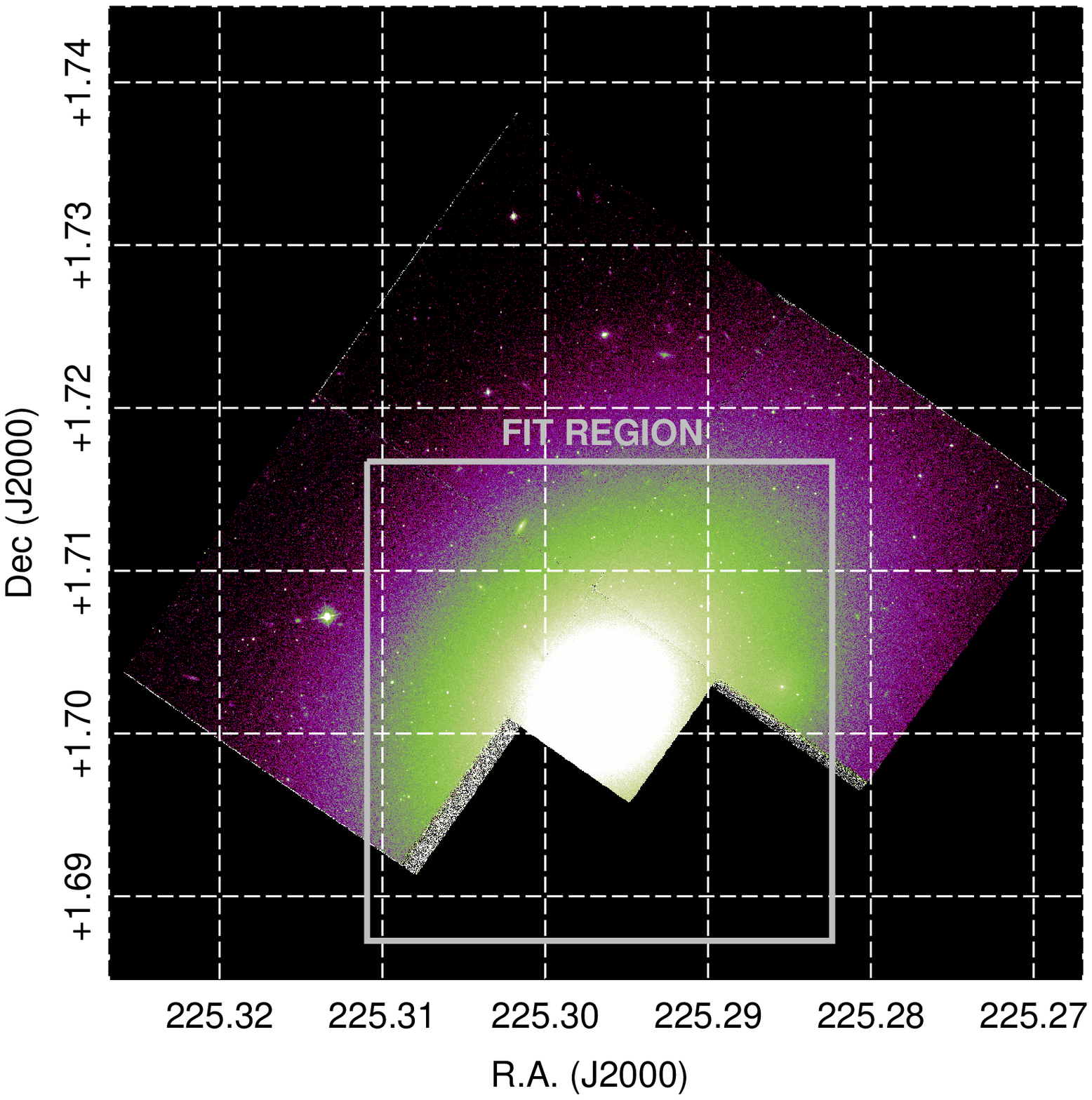}
   \put(20,75){\textcolor{white}{NGC 5813}}
  \end{overpic}
  \includegraphics[width=0.48\textwidth,angle=0]{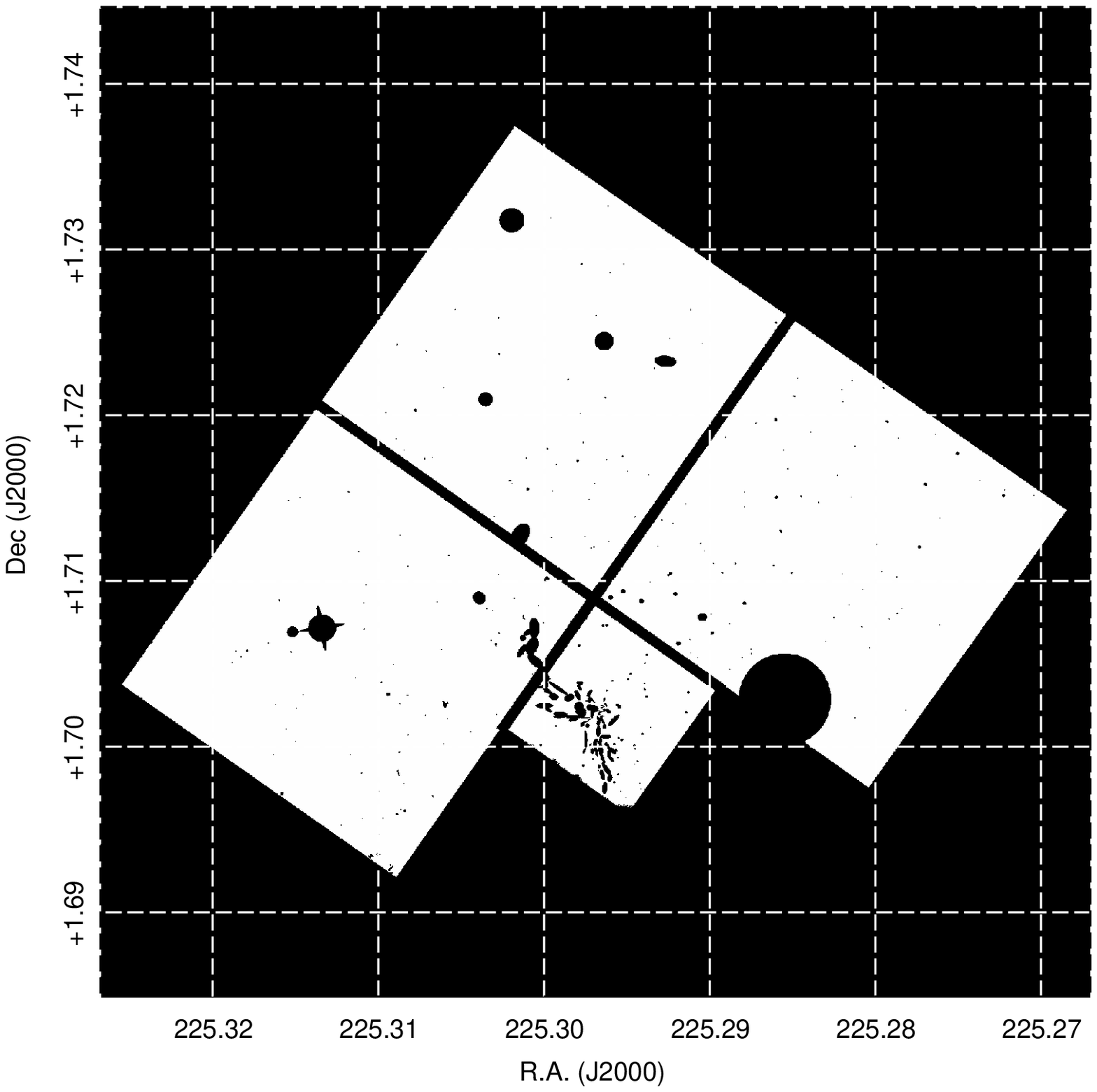}
 }
 \caption{
 HLA images (\emph{left}) and relevant masks (\emph{right}), for the sample
 galaxies \mbox{NGC 5557} (\emph{top}) and \mbox{NGC 5813} (\emph{bottom}).
 The grey box shows the extent of the fit (see discussion in
 \S\ref{Sky level estimation, sigma image, and fit extent}).
 All point-like sources and extended objects have been masked, as well as the
 bad pixels along the detectors seams and borders (see description in
 \S\ref{Masking}).
 In particular, a close inspection of the galaxy \mbox{NGC 5813} revealed a
 dust lane (not evident in the image with the displayed color scale), which was
 accurately masked by hand.
 \label{figure:mosaics_and_masks}
}
\end{figure*}

Due to the alleged HLA over-subtracted sky, we decided to \emph{limit} our fit extent to
roughly the point at which the galaxian brightness dropped to the estimated background
level.
For the same reason we did not include any sky component in the fit.
Figure \ref{figure:mosaics_and_masks} shows the HLA images for \mbox{NGC 5557} and
\mbox{NGC 5813}, and the region we fit.

\subsubsection[Masking]{Masking}
\label{Masking}

With our procedure, we aimed to mask any detector effect (e.g. hot pixels,
charge transfer trails, chip-to-chip seams, ghosts, etc.), as well as contaminating
objects.

In the first place, we dealt with the chip-to-chip seams by creating
masks covering the intersections between the different chips of the WFPC2 camera,
as well as the chip borders, where several ``deviant'' pixels can be found
(i.e., pixels with very low response or with high/erratic dark current).
As the relative chip positions change with time up to a shift of $\sim$5 pixels
\citep[][Table 5.9]{WFPC2}, it was not possible to automatize this process, and these
masks had to be adapted by hand.
In \S\ref{Assessing results and troubleshooting} we discuss the effects of
an unproper masking of such deviant pixels on the fit results.

Contaminating objects were masked by using a 2-step procedure based on
\textsc{SExtractor} \citep{SExtractor}.
A first, small (16 pixel) background mesh, \textsc{SExtractor} run was performed to
identify point-like sources, especially those embedded well within the innermost
galactocentric radii ($<$25\% of effective radius); for this purpose, the
deblending contrast was set to a particularly low value (0.0001), and filtering
turned off.
A second run with a bigger background mesh (64 pixels) aimed to identify
extended objects.
Using a mesh wider than the one we adopted resulted in missing small extended
sources near the target galaxy.

The segmentation maps obtained in this way (after excluding
the galaxy of interest) were combined to obtain the ``\textsc{SExtractor} mask''.
Finally dust, charge transfer trails, diffraction spikes, hot pixels, and 
contaminating sources were inspected and masked by hand, and the result was
combined with the \textsc{SExtractor} mask to get the final mask (right panels of
Figure \ref{figure:mosaics_and_masks}).

\subsubsection[PSF]{PSF}
\label{PSF}

Artificial PSFs  were produced using the TinyTim tool \citep{TinyTim}.
The HST PSF is diffraction limited, therefore its diffraction pattern expands
with increasing wavelength.
To account for this, the user has to suggest TinyTim a spectral profile
representative of the object under examination.
The software offers two choices of in-built stellar spectra: power-law and 
black body.
We chose to approximate the galaxian spectra with a black body distribution
characterized by a temperature of $\sim$4500~K, under the assumption that the
early type galaxies in our sample could be considered as a collection of coeval
K-type stars \citep[e.g.][]{lindsay}.
The PSF image size was chosen to be $\sim$30 times its FWHM in order to include
the extended wings of the PSF, and the convolution box was conservatively
set to several hundreds times the PSF FWHM, well beyond the core region.

\subsection[Fit]{Fit}
\label{Fit}

Initial guess fit parameters were chosen based on the results by \cite{dullo:2013b},
and by inspecting the 1D radial profiles along elliptical annulii of fixed
axis ratio and position angle generated via the IRAF.\emph{ellipse} task
\citep{ellipse}.
As noted earlier, the light profile of \mbox{NGC 5557} can be accurately described
by a single \corser{} component, while that of \mbox{NGC 5813} also requires a
disk component.

Fit results are presented in Table \ref{table:fit}.
The magnitudes have been zero-pointed to the STMAG system by using the relevant
STScI information reported in the images header (nominally, the \textsc{photflam}
keyword).
By construction, the STMAG magnitude for the F555W filter is very close
($\sim$0.01~mag brighter) to the Vegamag Cousin $V$-band magnitude
(see WFPC2 Instrument Handbook, version 10.0, \S{5.1}).
\GALFIT{} fit errors are not reported here, as the uncertainties derived from the
covariance matrix are usually underestimated \citep{GALFIT}.
We stress the fact that the values obtained for the reduced \chisq{} must be
considered with caution, as they are extremely dependent on the normalization
imposed to the sigma image by the evaluation of the background
(\S\ref{Sky level estimation, sigma image, and fit extent}).

\begin{figure*}
 \begin{minipage}{1.05\columnwidth}
  \includegraphics[width=\textwidth,angle=0]{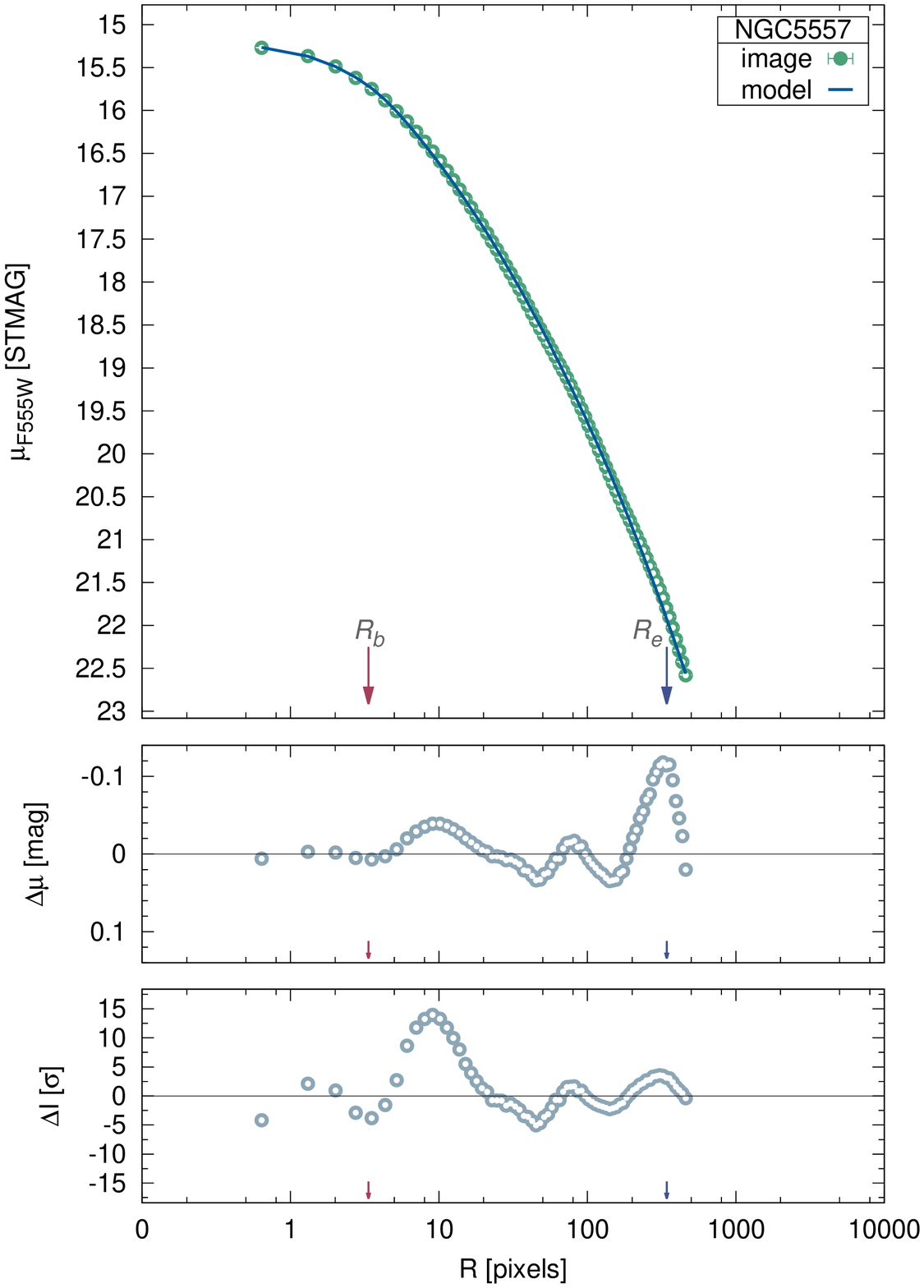}
 \end{minipage}
 \begin{minipage}{1.05\columnwidth}
  \includegraphics[width=\textwidth,angle=0]{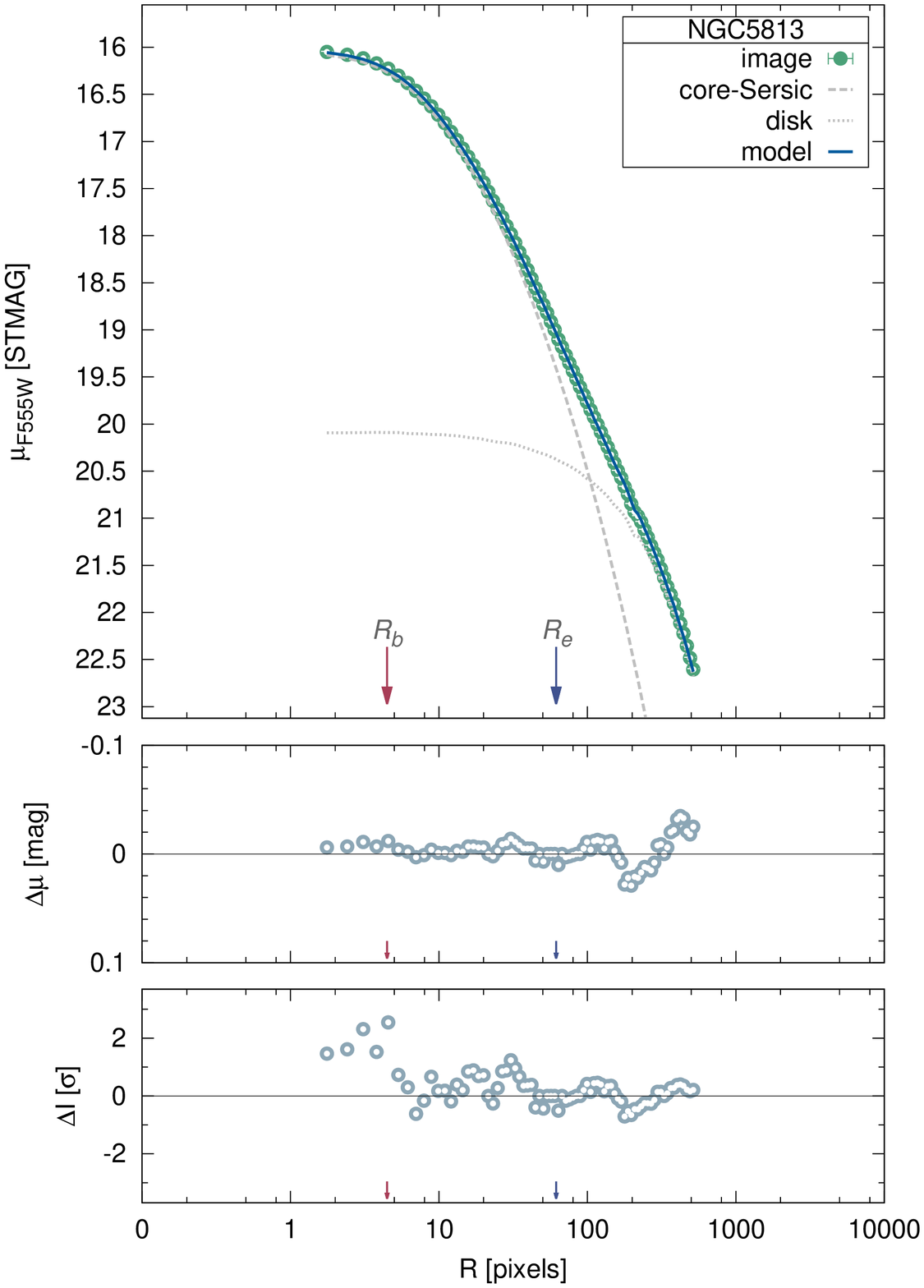}
 \end{minipage}
 \caption{
 Radial profiles of the galaxy light (\emph{green data points}), \GALFIT{}
 \emph{best-fit} [PSF-convolved] model/components (\emph{lines}), and residuals
 (\emph{grey data points}), for \mbox{NGC 5557} (\emph{left}) and \mbox{NGC 5813}    
 (\emph{right}).
 All data represent measurements along elliptical annulii of fixed
 axis ratio and position angle corresponding to the best-fit parameters, and
 centered at the center of the \corser{} component.
 The radial coordinate is to be considered as the semi-major axis value of each
 annulus; due to the presentation purposes of this work, radii are reported in
 units of pixels for a more readily comparison against the images.
 All the error bars for the image data are within the data symbols.
 We indicate with arrows the location of the break radius (\emph{red arrow}),
 and effective radius (\emph{blue arrow}) of the \corser{} model.
 The bottom panels represent the radial profiles of the residuals, expressed in
 terms of: (\emph{middle}) difference in surface brightness, and (\emph{bottom})
 residuals (in units of counts) over standard deviation as measured
 on the sigma image.
 \label{figure:radial_profiles}
}
\end{figure*}

Figure \ref{figure:radial_profiles} shows the radial profiles of the galaxy light
(obtained as mentioned above), of the \GALFIT{} \emph{best-fit} model, and of the
model residuals, for \mbox{NGC 5557} (\emph{left}) and \mbox{NGC 5813} (\emph{right}).

Inspecting the residuals (middle panels), we observe that the chosen models
provide a fairly good fit to the data, with residuals less than
0.1~mag/arcsec$^{2}$ over the extent of the fit: an accuracy comparable
to that of previous 1D analysis of the same objects
\citep[][their Figure A1]{dullo:2013b}.
Though, we notice a peculiar pattern of the residuals at the outermost radii,
corresponding to a steepening of the slope of the light profile: we attribute
this effect to possible sky over-subtraction
(see \ref{Sky level estimation, sigma image, and fit extent}).

In the bottom panels of Figure \ref{figure:radial_profiles}, we report the fit
residuals expressed in units of $\sigma$ (``weighted residuals'').
This visualization shows that, despite the \GALFIT{} Poissonian weighting scheme
is naturally biased towards the brightest - innermost - data,
the fit is actually constrained by the data at large radii, which show smaller
weighted residuals.
The reason is that the outer \Sersic{} part of model is sampled by a number of pixels
which is orders of magnitude larger than the few pixels of the power-law part,
hence it contributes more to the overall \chisq{}.
Indeed, we observe that the dispersion of the weighted residuals at the innermost
radii decreases when fitting a smaller area around the galaxy center.

The slight boxiness of the disk component of \mbox{NGC 5813} is to be attributed
to a distortion of the isophotes due partly to an anomalous intensity
discontinuity between the WF2 chip (which appears slightly brighter) and the other
chips, and partly to the dust ``waist'' which reduces the galaxy intensity along the
S-W to N-E direction (the dust pattern is evident in the image residuals; see Figure  
\ref{figure:resid}).

\begin{figure*}
 \begin{minipage}{1.05\columnwidth}
  \includegraphics[width=\textwidth,angle=0]{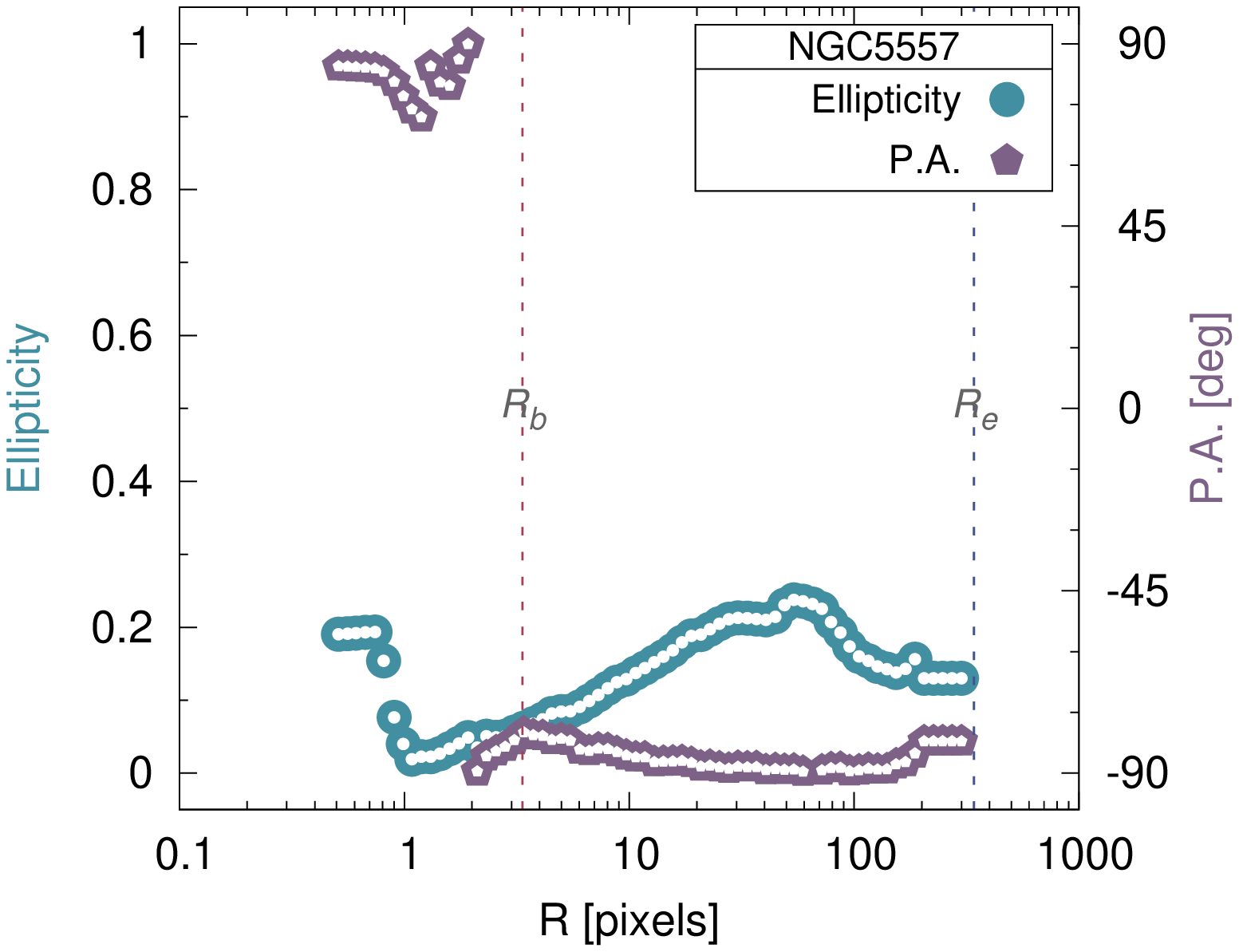}
 \end{minipage}
 \begin{minipage}{1.05\columnwidth}
  \includegraphics[width=\textwidth,angle=0]{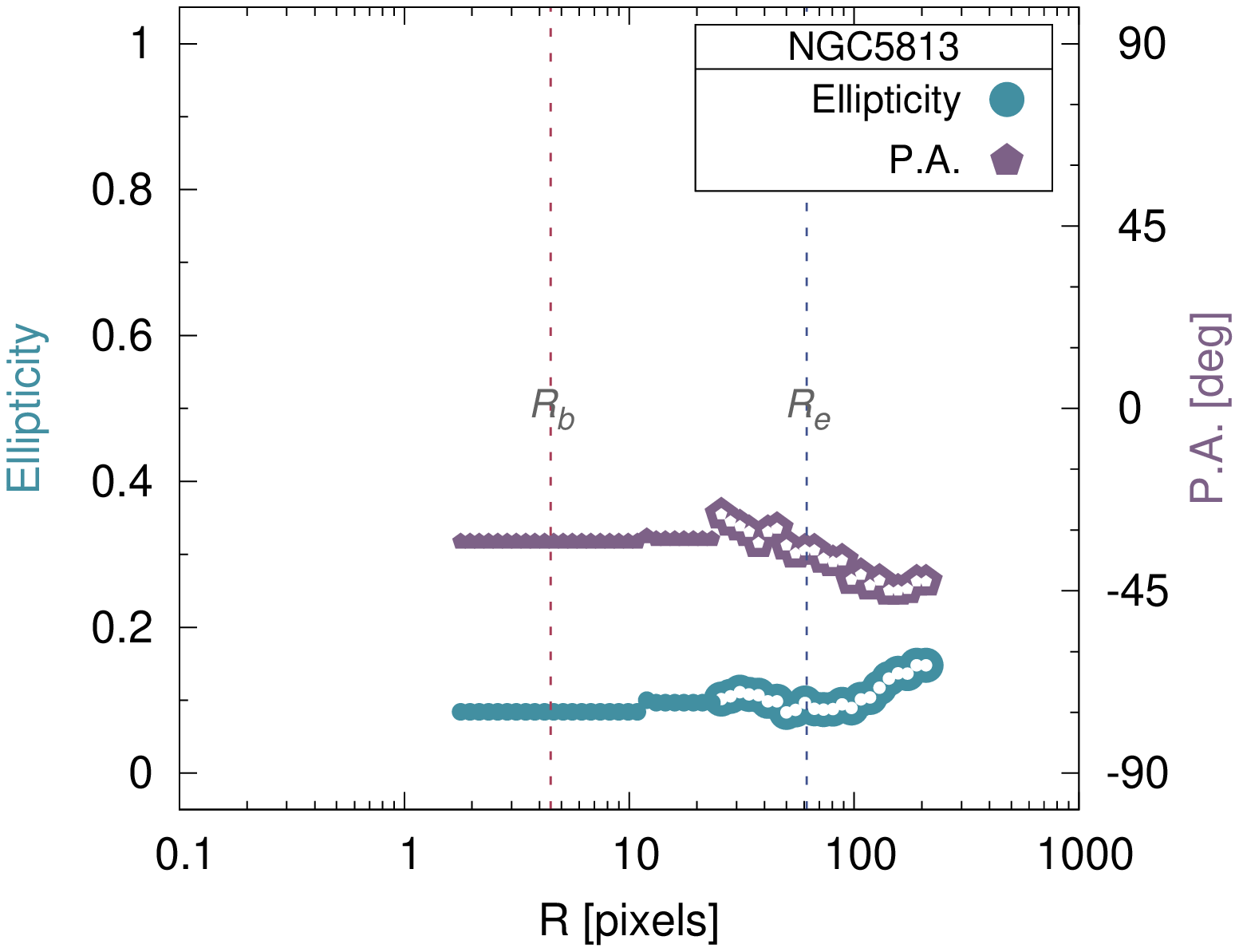}
 \end{minipage}
 \caption{
 Ellipticity (\emph{green circles}) and position angle (\emph{violet diamonds})
 profiles for \mbox{NGC 5557} (\emph{left}) and \mbox{NGC 5813} (\emph{right}),
 as measured along elliptical annulii by IRAF.\emph{ellipse}.
 Differently from the procedure followed to create
 Figure \ref{figure:radial_profiles}, the task was here allowed to fit the
 elliptical isophotes for ellipticity and position angle.
 The vertical dotted line represent the locations of the break radius
 (\emph{red line}) and of the effective radius (\emph{blue line}).
 The innermost regions of \mbox{NGC 5813} were severely affected by dust obscuration
 and were therefore heavily masked, hence the ellipticities and position angles
 calculated by IRAF.\emph{ellipse} up to R $\sim$ 25~pixels should be
 disregarded (those data are shown in the plot with a smaller marker size). 
 \label{figure:ellipse}
}
\end{figure*}

\begin{figure*}
 \begin{minipage}{1.05\columnwidth}
  \begin{overpic}[width=\textwidth]
   {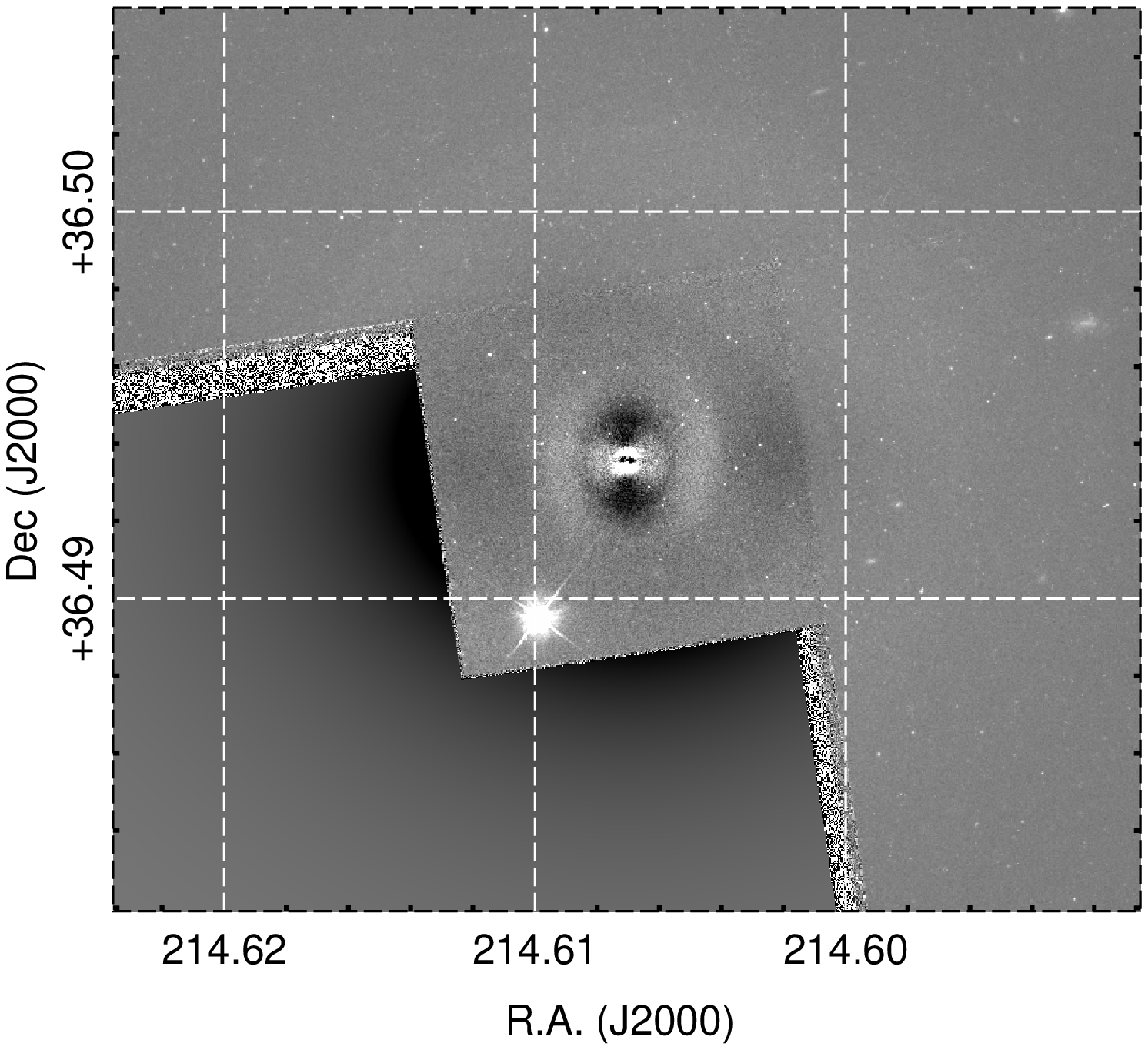}
   \put(40,90){\textcolor{black}{NGC 5557}}
  \end{overpic}
 \end{minipage}
 \begin{minipage}{1.05\columnwidth}
  \begin{overpic}[width=\textwidth]
   {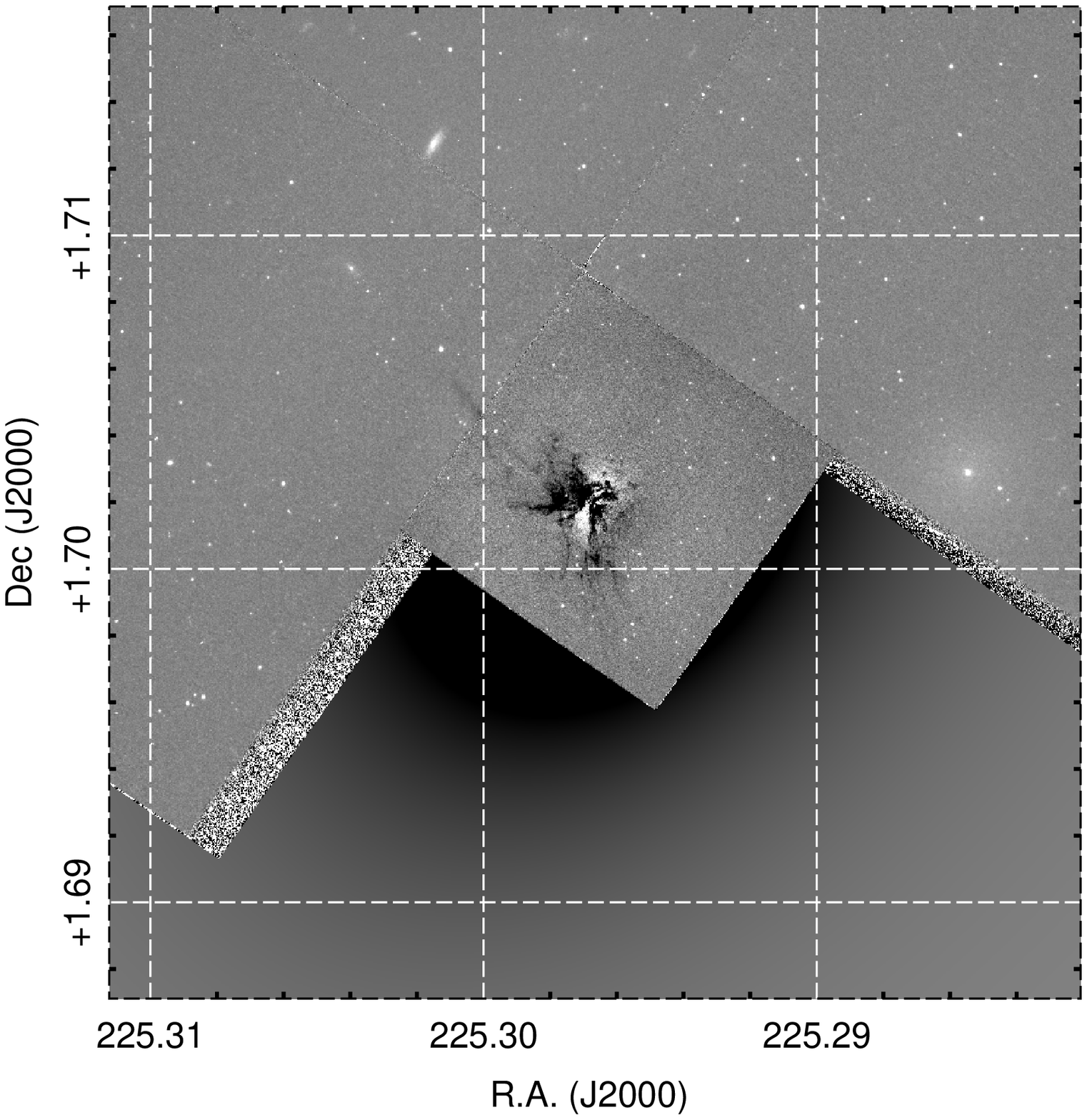}
   \put(40,90){\textcolor{black}{NGC 5813}}
  \end{overpic}
 \end{minipage}
 \caption{
  \GALFIT{} fit residuals for \mbox{NGC 5557} (\emph{left}), and \mbox{NGC 5813}
  (\emph{right}).
  The areas represented here correspond to the fit regions shown in Figure
  \ref{figure:mosaics_and_masks}.
  \label{figure:resid}
 }
\end{figure*}

In comparing our results against previous 1D studies,
the reader should bear in mind that (\emph{a}) we performed PSF-convolved fitting,
and that (\emph{b}) the photometry presented in Figure \ref{figure:radial_profiles}
was performed along elliptical annulii of \emph{fixed} axis ratio and position angle
corresponding to the best-fit parameters.
The reason for this choice is due to the fact that \GALFIT{} allows only for
\emph{one} value of axis ratio and position angle for each component of the
model.
Differences may arise with studies based on the IRAF.\emph{ellipse} task, when the
task is allowed to fit for ellipticity and position angle.
We performed such \emph{ellipse} analysis, and report in
Figure \ref{figure:ellipse} the trends of the ellipticities and position
angles of the fitted elliptical isophotes.
Indeed, it is the case that - at least for the galaxy \mbox{NGC 5557}
(\emph{lef} panel) - there is a significant variation of the ellipticity throughout
the galaxy extent.

Figure \ref{figure:resid} shows the residual images for the sub-section of the
HLA images (Figure \ref{figure:mosaics_and_masks}) over which the fit was performed.
The image for \mbox{NGC 5557} (\emph{left} panel)
shows a weak residual structure within the innermost radii ($<$500~pc), attributable
to the change in the ellipticity (Figure \ref{figure:ellipse}).
The residual image for \mbox{NGC 5813} (\emph{right} panel) clearly shows the trace
of its dust content (which was masked prior to fitting; see \S\ref{Masking}).

\subsection[Assessing results and troubleshooting]{Assessing results and troubleshooting}
\label{Assessing results and troubleshooting}

\subsubsection[Comparison against 1D results]{Comparison against 1D results}
\label{Comparison against 1D results}

In order to assess the reliability of our 2D analysis, we compared our
results against those of 1D \corser{} fits found in the literature.
Both \mbox{NGC 5557} and \mbox{NGC 5813} have been analysed based on WFPC2 images
in the F555W filter ($\sim${$V$}-band) by \cite{dullo:2013b}.
This work represents our main reference study, as we used data obtained with
the same instrument and filter setup.
\mbox{NGC 5557} was also investigated by \cite{trujillo:corser} in the $R$-band
(HST WFPC2/F702W), too far from our $V$-band to allow for a comparison without
using a statistically significant sample.
\mbox{NGC 5813} was studied by \cite{richings} in the
$I$-band (HST NICMOS/F814W).
However, \cite{richings} failed to fit the large scale disk component of
\mbox{NGC 5813}, deemed necessary in this work and by \cite{dullo:2013b}, hence
it was not possible to compare the results.
A comparison of the parameters obtained in this work against those by
\cite{dullo:2013b} are reported in Table \ref{table:comparison}.

We obtained a fairly good agreement, with the values reported by \cite{dullo:2013b}.
In their Figures 2, 3, and 4, these authors present a comparison between
their parameters $r_{b}$, \Sersic{} $n$, and $\mu_{b}$, and those obtained
by diverse 1D studies of core galaxies (including both Nuker and \corser{}
modeling): all of our values lay well within the scatter of these plots.

The only noticeable difference with \cite{dullo:2013b} is in the values of
the power-law slopes ($\gamma$), despite both our
(Figure \ref{figure:radial_profiles}) and their \citep[][their Figure A1]{dullo:2013b}
fits yielded to very low residuals inside the break radii.
For the photometry of the very inner part of the galaxies (R $<$ 1\arcsec),
\cite{dullo:2013b} used the data from \cite{lauer:2005}, which were obtained
through a PSF-deconvolution technique.
On the contrary, \GALFIT{} follows a PSF-fitting method working directly on
the convolved image.
We notice that, for both galaxies, the radial extent of the power-law part of
the \corser{} model is very close to the size of the PSF, as measured on the
HLA images ($\sim$2~pixels FWHM).
Therefore, it is plausible that the value of $\gamma$ is sensitive to the different
treatment of the PSF in the two methods.

In the case of \mbox{NGC 5813}, an additive issue is that the different inner
power-law slopes may be the reflection of a difference in the position of the
galaxian centers derived in the two works.
An offset center strongly affects the radial profile at the innermost radii,
and could even mimic a slope of opposite sign with respect to the true one.
In \GALFIT{}, the galaxian center is determined using information from the
\emph{whole} image, while the isophote fitting used in \cite{dullo:2013b}
uses the local pixels.
The inner regions of \mbox{NGC 5813} are severely affected by dust (see Figure
\ref{figure:mosaics_and_masks}), therefore it is possible that the different centering
methods and, most of all, the different choices for the dust mask in the two works
yielded to a different fit of the center of this galaxy.
Indeed, even in our \GALFIT{} fit, the value of $\gamma$ turned out to be very
sensitive to the masking choice.

Our values for the effective radii are only slightly different from those of
\cite{dullo:2013b} for both galaxies.
We ascribe the small discrepancy to the different radial extent of the fit:
\cite{dullo:2013b} were able to fit further, while we had to restrain our
fits due to the alleged over-subtraction of the sky
(see \S\ref{Sky level estimation, sigma image, and fit extent}).
Finally, we stress that the fits by \cite{dullo:2013b} were performed by
keeping the $\alpha$ parameter ``frozen'', hence part of the discrepancies
in the other parameters can be a reflection of this constraint.

\subsubsection[Troubleshooting]{Troubleshooting}
\label{Troubleshooting}

In this section we will present a short summary of the possible sources of
error we had to account for in the current analysis.

\medskip
\noindent 
\emph{Issues with HLA WFPC2 data.} 

In \S\ref{Sky level estimation, sigma image, and fit extent}, we extensively
presented the issues arising from the sky-subtraction procedure
performed by the HLA pipeline.
An over-subtraction of the sky contribution could easily mimic a truncation of
the external region of the disk, while an under-subtraction can be misinterpreted
as an extended disk.
In the specific cases examined here (\mbox{NGC 5557} and \mbox{NGC 5813}), we
inferred that the sky was over-subtracted.
This influenced the estimate of the effective radius ($r_{e}$).

In \S\ref{Masking} we discussed the importance of masking the bad
pixels (i.e., pixels with very low response or with high/erratic dark current)
at the borders of the detector, or along the seams between chips.
The effects of these deviant pixels can be easily recognized as they drive the
fit towards nonphysical position angles.
In \S\ref{Assessing results and troubleshooting}, we also discussed how dust can
influence the parameters of the inner part of the \corser{} model, and especially
$\gamma$.

\medskip
\noindent 
\emph{Notes on the \corser{} model in \GALFIT{}.} 

The parameter $\alpha$ is the one that required most attention in the
implementation of the \corser{} model.
In the first place, as discussed in \S\ref{Implementation of the corser model
in GALFIT}, high values of $\alpha$ can generate numerical overflows,
although we limited the onset of this problem by setting a hard upper boundary for
$\alpha$ (20).
We observed that in most of the cases in which the fit did not converged $\alpha$
was prone to be continuously reset to the upper boundary with each fit iteration.
In the case of small cores, or when the transition between the core and the
\Sersic{} region is very mild, the code may ``decide'' to bypass the core
by fitting low values of alpha, and essentially use only the \Sersic{} part
to fit the whole profile.
This has the immediate effect of coupling the $\gamma$ and $n$ parameters
(which regulate the inner and outer curvature), as noted also in
\cite{graham:corser}.

In order to achieve a robust fit, the break radius $r_b$ should be bigger than
PSF FWHM (and the PSF be Nyquist sampled).
This is related to the suggestion by \cite{trujillo:corser} that
$r_{b}$ should be beyond the radius corresponding to the second innermost data
point (of deconvolved light profiles) to obtain an unambiguous identification of
the core.

Finally, we recall that it is suggested to extend the fit up to the galactocentric
distance corresponding to the effective radius $r_{e}$ to obtain proper constraining
of the \Sersic{} part of the model \citep[see also][]{graham:corser},
although \citet{dullo:2012,dullo:2013b} have revealed that a careful modeling of
smaller radial extents can still recover the \Sersic{} parameters.

\section[Summary and conclusions]{Summary and conclusions}
\label{Summary and conclusions}

We introduced \CORSAIR{}: a publicly available, fully retro-compatible
modification of the 2D fitting software \GALFIT{} (v.3) which adds an
implementation of the \corser{} model.

We presented a practical application of the code through the analysis of Hubble
Legacy Archive (HLA; F555W filter) data of two ``secure'' \corser{} galaxies
\citep{trujillo:corser,richings,dullo:2013b}: \mbox{NGC 5557} and \mbox{NGC 5813}.
We showed that the code is robust with respect to multi-component fitting, even in
conditions of heavy dust obscuration. 
We compared our results against previous 1D analysis, and we find reasonable
consistency in the derived \corser{} parameters, considering the intrinsic
differences between the 1D and the 2D fitting methods. 
Finally, we reported a summary of warnings regarding the specific analysis of HLA
images, and on the usage of the \corser{} model.

The code and the analysis procedure described here have been developed to perform
the first coherent 2D analysis of a sample of \corser{} galaxies
\citep[the one of][]{dullo:2013b}, which will be presented in a
companion paper (Bonfini et al. 2014, in preparation).
As discussed in \S\ref{2D modeling of corser galaxies}, two dimensional fitting
proved to be superior to 1D approaches in multi-components analysis.
Moreover, \GALFIT{} fit is fully seeing-corrected, through the convolution
of the model with the PSF.
Therefore, our 2D analysis will yield complementary constraints on the missing
mass in depleted galaxy cores.

\acknowledgments

We thank A. Graham for his suggestions with the analysis, and input which helped
this project.
We are very grateful to C. Peng, who made possible the modification of \GALFIT,
and assisted on the debugging procedure.
We also wish to thank P. Erwin for the insights on the Imfit code.
Finally, we thank D. Gadotti for a valuable review of this document, which led
to a more accurate analysis of our data.
Based on observations made with the NASA/ESA Hubble Space Telescope, and obtained
from the Hubble Legacy Archive, which is a collaboration between the Space
Telescope Science Institute (STScI/NASA), the Space Telescope European
Coordinating Facility (ST-ECF/ESA) and the Canadian Astronomy Data Centre
(CADC/NRC/CSA).
The HST observations are associated with GO proposals 6587 (PI: Richstone, D.),
and 5454 (PI: Franx, M.).


\appendix

\section[Notes on sample galaxies]{Notes on sample galaxies}
\label{Notes on sample galaxies}

\subsection[NGC 5557]{NGC 5557}
\label{NGC 5557}

\mbox{NGC 5813} is a massive \citep[$M_{V}$ = -22.39~mag;][]{lauer:2007}, rounded
(see Table \ref{table:fit}) elliptical galaxy \citep[E1;][]{RC3}, and it has been
identified as a slow rotator by the ATLAS$^{3D}$ survey \citep{emsellem}.
These characteristics match the profile of the prototype \corser{} galaxies,
which are believed to be result of dry merger events (see \ref{corser galaxies}).
Contrarily, using deep MegaCam exposures, \cite{duc} found that NGC 5557 shows
clear evidence for extremely elongated tidal tails (for a total extent of 375~kpc),
and high fine-structure index \citep{schweizer}, which they attributed to a recent
(z$<$0.5) gas-rich major merger event.
Though, the authors acknowledge that this hypothesis can be reconciled with the
mild rotation of NGC 5557 only under the specific assumption that the merger
collision had a low impact parameter.

\subsection[NGC 5813]{NGC 5813}
\label{NGC 5813}

\mbox{NGC 5813} has been classified as elliptical galaxy (E1) in the 
Third Reference Catalogue \citep[RC3;][]{RC3}.
However, both \citet[their Appendix C]{trujillo:corser} and \cite{dullo:2013b}
found that the galaxy may be better described by a \corser{} + disk component,
hence suggesting it may be a S0 galaxy.
The results of our 2D analysis (\S\ref{Fit}) confirm that the galaxian light
profile is well fit by the aforementioned components.
It is worth mentioning, though, that the galaxy is believed to have a kinematically
decoupled core, as first found by \cite{efstathiou}, and interpreted by 
\cite{kormendy:NGC5813} as the result of a minor merger.
The kinematically decoupled core was also reported among the early SAURON results
by \cite{deZeeuw}.
Indeed, inspecting the ATLAS$^{3D}$ velocity maps \citep{ATLAS3D}, we observe
that the decoupled core extends up to $\sim$6\arcsec, beyond which the rotational
velocity suddenly drops.

\section[GALFIT-CORSAIR ``how to'']{G{\small ALFIT}-C{\small ORSAIR} ``how to''}
\label{CORSAIR how to}

\noindent
A pre-compiled version of \CORSAIR{} is available at the address:

\medskip
\url{http://astronomy.swin.edu.au/~pbonfini/galfit-corsair/}
\medskip

\noindent
along with an example input file, and usage notes.
The code is completely retro-compatible with \GALFIT{} v.3, but, at the moment
this document is written, the \corser{} model has \emph{not} been tested yet for
the azimuthal/truncation functions, and it is not possible to apply constraints on
its parameters $r_{b}$, $\alpha$, $\gamma$, $r_{e}$, and $n$.
Further updates will be notified on the website.

\noindent
The definition of the \corser{} model follows the standards of the \GALFIT{} input
parameter file:

\medskip
\begin{minipage}{\columnwidth}
 \begin{alltt}
# core-Sersic function
 
 0) corser         # object type
 1) 123 456   1 1  # position x, y        [pixel]			       
 3) -10.0  	  1	   #    mu(Rb)            [surface brightness mag. at Rb] 
 4) 4.0       1    #     R_b              [pixel]			       
 5) 5.0       1    #     alpha  					  
 6) 0.1       1    #     gamma  					  
 7) 123       1    #     R_e              [pixel]			       
 8) 4         1    # Sersic exponent (deVauc=4, expdisk=1)	       
 9) 0.90      1    # axis ratio (b/a)					       
10) 90        1    # position angle (PA)  [Degrees: Up=0, Left=90]     
C0) 1e-05     1    # diskyness(-)/boxyness(+)			       
 Z) 0              # leave in [1] or subtract [0] this comp from data? 
 \end{alltt}
\end{minipage}

\noindent
NOTE 1: The \corser{} flux normalization (parameter \#3) can \emph{only} be defined at
the model's native size parameter ($R_{b}$), therefore the other normalizations
\citep[as per][\S{6.1}]{GALFIT}, will not be recognized.

\medskip
\noindent
NOTE 2: As for the other models in \GALFIT{}, the diskyness/boxyness parameter (C0)
shall be always set to an initial value different than 0 if this parameter is fit,
or else the software will crash.

\medskip
\noindent
Please report any bug to: \texttt{pbonfini@swin.edu.au}.





\begin{deluxetable}{ccccccc}
 \tabletypesize{\small}
 \tablecaption{Sample galaxies: image characteristics \label{table:sample_images}}
 \tablewidth{0pt}
 \tablehead{
  \colhead{Target}        &
  \colhead{RA  (J2000)}   &
  \colhead{Dec (J2000)}   &
  \colhead{$D$}           &
  \colhead{Camera/Filter} &
  \colhead{N$_{exp}$}     &
  \colhead{Exp. time} 
  \\
  \colhead{}           &
  \colhead{[hh:mm:ss]} &
  \colhead{[dd:mm:ss]} &
  \colhead{[Mpc]}      &
  \colhead{}           &
  \colhead{}           &
  \colhead{[sec]} 
  \\
  \colhead{{\tiny (1)}} &
  \colhead{{\tiny (2)}} &     
  \colhead{{\tiny (3)}} &
  \colhead{{\tiny (4)}} &
  \colhead{{\tiny (5)}} &
  \colhead{{\tiny (6)}} &
  \colhead{{\tiny (7)}}
 }
 \startdata
NGC 5557 & 14:18:25.7 & +36:29:37 & 47.9 & WFPC2/F555W & 6 & 1400 \\
NGC 5813 & 15:01:11.2 & +01:42:07 & 32.2 & WFPC2/F555W & 2 & 1000 \\
 \enddata
 \tablecomments{
  Details of the images for the sample galaxies.
  The final HLA products used in the current work are sky-subtracted mosaics of the
  observations for each galaxy.
  The number of combined images and exposure time are relevant to the statistics used
  by \GALFIT{} to perform the fit (see \S\ref{Data preparation and fitting})
  \\
  $^{(1)}$ Target name.
  $^{(2,3)}$ Target coordinates from NED (J2000).
  $^{(4)}$ Target distance: from \cite{cantiello} for \mbox{NGC 5557}, and
           from \cite{tonry} for \mbox{NGC 5813}.
  $^{(5)}$ HST camera and filter.
  $^{(6)}$ Number of exposures combined to obtain the HLA product.
  $^{(7)}$ Total exposure time.
 }
\end{deluxetable}

\begin{deluxetable}{cccccccccccccr}
 \tabletypesize{\small}
 \tablecaption{Galfit fit results \label{table:fit}}
 \tablewidth{0pt}
 \tablehead{
  \colhead{Target}                     &
  \colhead{Component}                  &
  \colhead{$\mu_{b, F555W}^{\dagger}$} & 
  \colhead{$m_{F555W}^{\dagger}$}      & 
  \colhead{$r_{b}$}                    &
  \colhead{$\alpha$}                   &
  \colhead{$\gamma$}                   &
  \colhead{$r_{e}$}                    &
  \colhead{$r_{s}$}                    &
  \colhead{$n$}                        &
  \colhead{$b$/$a$}                    &
  \colhead{P.A.}                       &
  \colhead{$C$}                        &
  \colhead{$\chi_{\nu}^{2}$} 
  \\
  \addlinespace 
  \colhead{}                    &
  \colhead{}                    &
  \colhead{[mag/\arcsec$^{2}$]} &
  \colhead{[mag]}               &
  \colhead{[pc]}                &
  \colhead{}                    &
  \colhead{}                    &
  \colhead{[~\arcsec~]}         &
  \colhead{[~\arcsec~]}         &
  \colhead{}                    &
  \colhead{}                    &
  \colhead{[deg]}               &
  \colhead{}                    &
  \colhead{}           
  \\
  \colhead{{\tiny (1)}}  &
  \colhead{{\tiny (2)}}  &     
  \colhead{{\tiny (3)}}  &
  \colhead{{\tiny (4)}}  &
  \colhead{{\tiny (5)}}  &
  \colhead{{\tiny (6)}}  &
  \colhead{{\tiny (7)}}  &
  \colhead{{\tiny (8)}}  &
  \colhead{{\tiny (9)}}  &
  \colhead{{\tiny (10)}} &
  \colhead{{\tiny (11)}} &
  \colhead{{\tiny (12)}} &
  \colhead{{\tiny (13)}} &
  \colhead{{\tiny (14)}}
 }
 \startdata
NGC 5557 & \corser{} & 15.62 & -     & 77.8 & 10.1 & 0.38 & 34.2 & -    & 5.0 & 0.82 &  90.5 & -0.08 & 17.3 \\
 \addlinespace 
NGC 5813 & \corser{} & 16.20 & -     & 69.8 & 2.5  & 0.01 &  6.2 & -    & 2.7 & 0.94 & 149.4 & -0.03 & 4.7 \\
NGC 5813 & disk      & -     & 11.50 & -    & -   & -    & -    & 24.4 & -   & 0.69 & 134.8 & +0.17 & -   \\
 \enddata
 \tablecomments{
  Results of the \GALFIT{} \corser{} analysis.
  \\
  $^{(1)}$  Target name.
  $^{(2)}$  Model component.
  $^{(3)}$  Surface brightness at break radius.
  $^{(4)}$  Total apparent magnitude.
  $^{(5)}$  Break radius for the \corser{} model.
            Lengths have been calculated assuming linear approximation of the
            angular separations.
  $^{(6)}$  Alpha parameter for the \corser{} model.
  $^{(7)}$  Inner power-law index for the \corser{} model.
  $^{(8)}$  Effective radius.
  $^{(9)}$  Disk scale length.
  $^{(10)}$ \Sersic{} index.
  $^{(11)}$ Axis ratio.
  $^{(12)}$ Position angle (east: P.A. = 0).
  $^{(13)}$ Boxyness of generalized ellipse (boxy isophote: $C$ $>$ 0).
  $^{(14)}$ Reduced \chisq{} for the whole model.
 }
 \tablenotetext{\dagger}{
  Values refer to the STMAG system; the STMAG magnitude for the F555W
  filter is very close (0.01~mag brighter) to the Cousin $V$-band magnitude.
 }
\end{deluxetable}

\begin{deluxetable}{clcccccc}
 \tabletypesize{\small}
 \tablecaption{Comparison of core-Sersic parameters against literature\label{table:comparison}}
 \tablewidth{0pt}
 \tablehead{
  \colhead{Target}            &
  \colhead{Source}            &
  \colhead{$\mu_{b}^{F555W}$} & 
  \colhead{$r_{b}$}           &
  \colhead{$\alpha$}          &
  \colhead{$\gamma$}          &
  \colhead{$r_{e}$}           &
  \colhead{$n$}               
  \\
  \addlinespace 
  \colhead{}              &
  \colhead{}              &
  \colhead{[STMAG]}       &
  \colhead{[arcsec]}      &
  \colhead{}              &
  \colhead{}              &
  \colhead{[arcsec]}      &
  \colhead{}              
  \\
  \colhead{{\tiny (1)}}  &
  \colhead{{\tiny (2)}}  &     
  \colhead{{\tiny (3)}}  &
  \colhead{{\tiny (4)}}  &
  \colhead{{\tiny (5)}}  &
  \colhead{{\tiny (6)}}  &
  \colhead{{\tiny (7)}}  &
  \colhead{{\tiny (8)}} 
 }
 \startdata
NGC 5557 & This work              & 15.62 & 0.34 & 10.1 &  0.38 & 34.2 & 5.0 \\
NGC 5557 & \cite{dullo:2013b}     & 15.47 & 0.23 & |5|  &  0.19 & 30.2 & 4.6 \\
 \addlinespace 
 \addlinespace 
NGC 5813 & This work              & 16.21 & 0.45 & 2.5  &  0.01 &  6.2 & 2.7 \\
NGC 5813 & \cite{dullo:2013b}     & 16.12 & 0.35 & |2|  & -0.10 &  7.1 & 2.8 \\
 \enddata
 \tablecomments{
  Comparison of the \corser{} parameters obtained through our 2D \GALFIT{} analysis
  against the work by \cite{dullo:2013b} based on 1D analysis in the same
  instrument/band (WFPC2/F555W).
  The $\alpha$ parameter in \cite{dullo:2013b} has been held fixed in the fit.
  \\
  $^{(1)}$  Target name.
  $^{(2)}$  reference study.
  $^{(3)}$  Break radius surface brightness.
            \cite{dullo:2013b} obtained their $V$-band magnitudes from the STMAG
            magnitudes by applying the -0.01~mag shift prescribed by the HST WFPC2
            Photometry Cookbook (Dullo, private communication). 
            Their values have been converted back to the STMAG system, in this table.
  $^{(4)}$  Break radius for the \corser{} model.
  $^{(5)}$  Alpha parameter for the \corser{} model.
  $^{(6)}$  Inner power-law index for the \corser{} model.
  $^{(7)}$  Effective radius of the \corser{} model.
  $^{(8)}$  \Sersic{} index.
 }
\end{deluxetable}



\end{document}